\documentclass [preprint,showpacs,nofootinbib]{revtex4}
\usepackage{bbm}
\usepackage{verbatim}
\usepackage{slashed}
\usepackage{epsfig}
\usepackage{feynmf}
\usepackage{graphicx}
\usepackage{amsmath}
\usepackage{mathrsfs}
\usepackage{bm}

\begin{document}


\title{Relativistic correction to $e^{+}e^{-}\to J/\psi+gg$
at $B$ factories and constraint on color-octet matrix elements}

\author{Zhi-Guo He\footnote{Present address: Departament d'Estructura i
Constituents de la Mat\`eria and Institut de Ci\`encies del
Cosmos,Universitat de Barcelona, Diagonal, 647, E-08028 Barcelona,
Catalonia, Spain. }}
\author{Ying Fan}
\author{Kuang-Ta Chao}

\affiliation{Department of Physics and State Key Laboratory of
Nuclear Physics and Technology, Peking University, Beijing 100871,
China}


\begin{abstract}\vspace{5mm}
We calculate the relativistic correction to $J/\psi$ production in
the color-singlet process $e^{+}e^{-}\to J/\psi+gg$ at $B$-factories.
We employ the nonrelativistic QCD  factorization approach,
where the short-distance coefficients are calculated perturbatively
and the long-distance matrix elements are extracted from the decays
of $J/\psi$ into $e^{+}e^{-}$ and light hadrons. We find that the
$O(v^2)$ relativistic correction can enhance the cross section by a
factor of 20-30\%, comparable to the enhancement due to the
$O(\alpha_s)$ radiative correction obtained earlier. Combining the
relativistic correction with the QCD radiative correction, we find
that the color-singlet contribution to $e^{+}e^{-}\to J/\psi+gg$ can
saturate the latest observed cross section $\sigma(e^{+}e^{-}\to
J/\psi+X_{\mathrm{non-c\bar{c}}})=0.43 \pm0.09\pm0.09$ pb by Belle,
thus leaving little room to the color-octet contributions. This
gives a very stringent constraint on the color-octet contribution,
and may imply that the values of color-octet matrix elements are
much smaller than expected earlier by using the naive velocity
scaling rules or extracted from fitting experimental data with the
leading-order calculations.
\end{abstract}


\maketitle
\section{Introduction}

The nonrelativistic QCD (NRQCD) effective field theory, introduced
in 1995 by Bodwin, Braaten, and Lepage\cite{Bodwin:1994jh}, has been
widely accepted as a rigorous approach to study the production and
decay of heavy quarkonium, the bound state of heavy quark $Q$ and
antiquark $\bar{Q}$ pair (see Ref.\cite{Brambilla:2004wf} for a
review). In the framework of NRQCD the production of heavy
quarkonium is factorized into two parts, the short-distance part and
the long-distance part. In the short-distance part, the $Q\bar{Q}$
pair is created in certain $J^{\mathrm{PC}}$ and color states, which can be
calculated perturbatively through the expansion of QCD coupling
constant $\alpha_{s}$. The long-distance part describes the
evolution of the $Q\bar{Q}$ pair into the physical hadron states
through the emission of soft gluons with the corresponding universal
nonperturbative matrix elements, which are weighted by powers of
the relative velocity $v_{_Q}$ of heavy quarks in the meson rest
frame. One intrinsic character of NRQCD is the inclusion of the
effect of $Q\bar{Q}$ pair in a color-octet state, $i.e.$ the
color-octet mechanism. Since in the $e^{+}e^{-}$ collision, the
structure of the parton involved is simpler and the signals can be
prominent, it is a good place to study the heavy quarkonium
production and test the color-octet mechanism.

In recent years, the cross sections of inclusive $J/\psi$ production
in $e^{+}e^{-}$ annihilation at $\sqrt{s}=10.6\mathrm{GeV}$ have
been reported by BaBar\cite{Aubert:2001pd}, Belle\cite{Abe:2001za},
and CLEO\cite{Briere:2004ug} collaborations with the results
\begin{subequations}
\begin{eqnarray}
\sigma(e^{+}e^{-}\to J/\psi+X)=2.5\pm{0.21}\pm{0.21}\mathrm{pb},
\end{eqnarray}
\begin{eqnarray}
\sigma(e^{+}e^{-}\to J/\psi+X)=1.47\pm{0.10}\pm{0.13}\mathrm{pb},
\end{eqnarray}
\begin{eqnarray}
\sigma(e^{+}e^{-}\to J/\psi+X)=1.9\pm{0.2}\mathrm{pb},
\end{eqnarray}
\end{subequations}
respectively. If the $J/\psi$'s momentum $p^{\ast}_{\psi}$ is
restricted to $p^{\ast}_{\psi}>2\mathrm{GeV/c}$, the results of
BaBar\cite{Aubert:2001pd} and Belle\cite{Abe:2001za} become
\begin{subequations}
\begin{eqnarray}
\sigma(e^{+}e^{-}\to
J/\psi+X)\Big{|}_{p^{\ast}_{\psi}>2\mathrm{GeV/c}}=
1.87\pm{0.10}\pm{0.15}\mathrm{pb},
\end{eqnarray}
\begin{eqnarray}
\sigma(e^{+}e^{-}\to
J/\psi+X)\Big{|}_{p^{\ast}_{\psi}>2\mathrm{GeV/c}}=
1.05\pm{0.04}\pm{0.09}\mathrm{pb}.
\end{eqnarray}
\end{subequations}

It can be seen that these experimental measurements are not entirely
consistent with each other. The two $B$-factories also measured the
$J/\psi$ momentum distributions and observed similar shapes. Despite
of the disagreement among different experimental measurements,
theoretically the inclusive $J/\psi$ production has been extensively
investigated within the color-singlet
model\cite{Driesen:1993us,Cho:1996cg,Yuan:1996ep,Baek:1998yf} and
the color-octet mechanism in
NRQCD\cite{Braaten:1995ez,Yuan:1996ep,Wang:2003fw}. In NRQCD, one
important contribution to the inclusive $J/\psi$ production in
$e^{+}e^{-}$ annihilation at $\sqrt{s}=10.6\;\mathrm{GeV}$ comes
from the color-octet process $e^{+}e^{-}\to
J/\psi[^1S_0^{(8)},^3P_{J}^{(8)}]+g$
\cite{Braaten:1995ez,Yuan:1996ep}. But its predictions of the
$J/\psi$ enhancement near the kinematics end point region is not
observed. After applying the soft-collinear effective theory
(SCET)\cite{Fleming:2003gt}, the shape of $J/\psi$ momentum
distribution can be softened. However this depends
phenomenologically on a nonperturbative shape function.

Experimentally, the $e^{+}e^{-}\to J/\psi+X$ process can be divided
into two parts: the $e^{+}e^{-}\to J/\psi+c\bar{c}+X$ part and the
$e^{+}e^{-}\to J/\psi+X_{\mathrm{non-c\bar{c}}}$ part. The Belle
collaboration finds the ratio\cite{Abe:2002rb}
\begin{equation}
R_{\mathrm{c\bar{c}}}=\frac{\sigma(e^{+}e^{-}\to J/\psi+c\bar{c}+X)}
{\sigma (e^{+}e^{-}\to J/\psi+X)}=0.59^{+0.15}_{-0.13}\pm{0.12}.
\end{equation}
which corresponds to $\sigma(e^{+}e^{-}\to
J/\psi+c\bar{c}+X)=0.87^{+0.21}_{-0.19}\pm{0.17} \mathrm{pb}$. The
latest measurement of $J/\psi$ production in association with a
$c\bar{c}$ pair carried out by Belle gives\cite{:2009nj}
\begin{equation}
\sigma(e^{+}e^{-}\to J/\psi+c\bar{c}+X)=0.74\pm
0.08^{+0.09}_{-0.08}\mathrm{pb}.
\end{equation}
The experimental results are more than 5 times larger than
leading-order (LO) NRQCD predictions\cite{Cho:1996cg,Baek:1998yf,Liu:2003jj}.
This large discrepancy could be resolved by including the NLO QCD
corrections and the feed-down of higher excited
states\cite{Zhang:2006ay,Gong:2009ng}. Belle also analyzed the
$e^{+}e^{-}\to J/\psi+X_{\mathrm{non-c\bar{c}}}$ process and
obtained\cite{:2009nj}
\begin{equation}\label{noncc}
\sigma(e^{+}e^{-}\to J/\psi+X_{\mathrm{non-c\bar{c}}})=0.43
\pm0.09\pm0.09\;\mathrm{pb}.
\end{equation}

At LO in $\alpha_s$ the $\mathrm{non-c\bar{c}}$ process
via color-singlet channels only includes $e^{+}e^{-}\to J/\psi+gg$.
The theoretical prediction at LO in $\alpha_s$ and $v^2$ is about
$0.2\mathrm{pb}$\cite{Yuan:1996ep}, and recent
works\cite{Ma:2008gq,Gong:2009kp} find that the K-factor of the NLO
correction is about 1.2-1.3. In charmonium system the relative
velocity $v_c$ of a $c\bar{c}$ pair in the $J/\psi$ rest frame is
not small. The value of $v_c^2$ is about $0.3$, which is close to
the size of $\alpha_s(m_c)$. So the relativistic effect may also be
important. Our previous work\cite{He:2007te} shows that in the
$e^{+}e^{-}\to J/\psi+\eta_c$ exclusive process, the relativistic
correction is indeed important in resolving the more than one order
of magnitude discrepancies between experimental
data\cite{Pakhlov:2004au,Aubert:2005tj} and LO NRQCD
predictions\cite{Braaten:2002fi}. We find when combining the
relativistic correction together with the NLO QCD
corrections\cite{Zhang:2005cha} the conflict between experimental
measurement and theoretical prediction is almost resolved (see also
Ref.~\cite{Bodwin:2006ke} for a similar result). However, in the
$e^{+}e^{-}\to J/\psi+c\bar{c}$ process, we find the relativistic
correction is very small and may be ignored. And the recent work
shows\cite{Fan:2009zq} that in the $p\bar{p}\to J/\psi+X$ process at
the Tevatron the relativistic corrections can also be neglected.
Then it is necessary to investigate the relativistic correction to
the $e^{+}e^{-}\to J/\psi+gg$ process and clarify how large the
relativistic correction is and whether it is positive or negative.
In this work we will deal with this problem within the framework of
NRQCD factorization approach, and in particular, we will examine the
effect of the relativistic correction on the constraint on
color-octet matrix elements. The rest of this paper is organized as
follows. In Sec.II, we will introduce the NRQCD factorization
formula and describe how we calculate the short-distance
coefficients and determine the long-distance matrix elements. We
will present our calculations and show our result in Sec.III.
Discussions and a summary will be given in the last section.

\section{The NRQCD Factorization Formula}

According to NRQCD\cite{Bodwin:1994jh} effective theory, up to
$\mathcal{O}(v^2)$ the inclusive $J/\psi$ production rate can be
expressed as
\begin{equation}
\sigma(e^{+}e^{-}\to
J/\psi+X)=\frac{F_{1}({}^{3}S_{1})}{m_{c}^2}\langle
0|\mathcal{O}_{1}^{\psi}({}^{3}S_{1})|0\rangle+\frac{G_{1}({}^{3}S_{1})}{m_{c}^4}\langle
0|\mathcal{P}_{1}^{\psi}({}^{3}S_{1})|0\rangle+O(v^{4}\sigma).
\end{equation}
where
\begin{equation}
\mathcal{O}_{1}^{\psi}({}^{3}S_{1})=\chi^{\dagger}\sigma^{i}
\psi(a_{\psi}^{\dagger}a_{\psi})\psi^{\dagger}\sigma^{i}\chi
\end{equation}
and
\begin{equation}
\mathcal{P}_{1}^{\psi}({}^{3}S_{1})=\frac{1}{2}[\chi^{\dagger}\sigma^{i}\psi(a_{\psi}^{\dagger}a_{\psi})
\psi^{\dagger}\sigma^{i}(-\frac{i}{2}\overleftrightarrow{\mathbf{D}})^2\chi+
\chi^{\dagger}\sigma^{i}(-\frac{i}{2}\overleftrightarrow{\mathbf{D}})^2\psi(a_{\psi}^{\dagger}a_{\psi})
\psi^{\dagger}\sigma^{i}\chi]
\end{equation}
are the four-fermion operators with dimensions six and eight, respectively.
$F_{1}({}^{3}S_{1})$ and $G_{1}({}^{3}S_{1})$ are the corresponding
short-distance coefficients. The short-distance coefficients can be
obtained perturbatively through the matching condition
\begin{equation}\label{mat}
\sigma(Q\overline{Q})\Big{|}_{\textrm{pert
QCD}}=\sum_{n}\frac{F_{n}(\Lambda)}{m_c^{d_{n}-4}}\langle0|\mathcal{O}_{n}^{Q\overline{Q}}
(\Lambda)|0\rangle\Big{|}_{\textrm{pert NRQCD}}
\end{equation}
The long-distance matrix elements characterized by the velocity
$v_c$ can be estimated by lattice calculations or phenomenological
models, or determined by fitting experimental data.

In this work, the covariant spinor projection
method\cite{Kuhn:1979bb} is adopted to evaluate the left-hand side
of Eq. [\ref{mat}]. In this method, the Dirac spinor product
$v(P/2-q)\overline{u}(P/2+q)$ is projected onto a certain
$^{(2S+1)}L_{J}$ state in a Lorentz covariant form (see,
e.g.,\cite{Keung:1982jb,Bodwin:2002hg}), which makes the
short-distance coefficients evaluated directly. In the $J/\psi$
(with $S=1$) case, the expression of the spinor production
projection in the meson rest frame up to all orders of $v_c^2$
is\cite{Bodwin:2002hg}
\begin{eqnarray}\label{eq10a}
\sum_{\lambda_{1}\lambda_{2}}v(\mathbf{-q},\lambda_{2})
\overline{u}(\mathbf{q},\lambda_{1})\langle\frac{1}{2},\lambda_{1};
\frac{1}{2},\lambda_{2}|1,\epsilon\rangle=
 \nonumber \\
\frac{1}{\sqrt{2}}(E+m)(1-\frac{\mbox{\boldmath$\alpha$}\cdot\mathbf{q}}{E+m})
{\mbox{\boldmath$\alpha$}}\cdot{\mbox{\boldmath$\epsilon$}}
\frac{1+\gamma_{0}}{2}(1+\frac{\mbox{\boldmath$\alpha$}\cdot\mathbf{q}}{E+m})\gamma_{0}.
\end{eqnarray}
And in an arbitrary frame, it becomes
\begin{eqnarray}\label{eq10b}
\sum_{\lambda_{1}\lambda_{2}}v(q,\lambda_{2})
\overline{u}(q,\lambda_{1})\langle\frac{1}{2},\lambda_{1};
\frac{1}{2},\lambda_{2}|1,\epsilon\rangle=
 \nonumber \\
-\frac{1}{2\sqrt{2}(E+m)}(\frac{1}{2}\slashed{P}-
\slashed{q}-m)\slashed{\epsilon}\frac{\slashed{P}+2E}{2E}
(\frac{1}{2} \slashed{P}+\slashed{q}+m).
\end{eqnarray}
Here the normalization of the Dirac spinors is
$\bar{u}u=-\bar{v}v=2m_c$, and the relations between momenta of
quark and antiquark in an arbitrary frame and in the meson rest
frame are given by\cite{Braaten:1996jt}
\begin{subequations}
\begin{equation}
\frac{1}{2}P+q=L(\frac{1}{2}P_{r}+\mathbf{q}),
\end{equation}
\begin{equation}
\frac{1}{2}P-q=L(\frac{1}{2}P_{r}-\mathbf{q}),
\end{equation}
\end{subequations}
where $P_{r}^{\mu}=(2E_{q},\mathbf{0})$,
$E_{q}=\sqrt{m^{2}+\mathbf{q}^{2}}$, and $2\mathbf{q}$ is the
relative momentum between two quarks in the meson rest frame.
$L_{\mu}^{v}$ is the boost tensor from the meson rest frame to an
arbitrary frame.

Then the cross section for $e^{+}e^{-}\rightarrow J/\psi+gg$ up to
next-to-leading order in $v^{2}$ can be expressed as
\begin{align}\label{final}
\sigma(e^{+}e^{-}\rightarrow J/\psi+gg)=
\frac{\langle0|\mathcal{O}_{1}^{\psi}({}^{3}S_{1})|0\rangle}{3}
\frac{1}{2s}\int\overline{N_{0}}\;d\Phi_3+
\frac{\langle0|\mathcal{P}_{1}^{\psi}({}^{3}S_{1})|0\rangle}{3}
\frac{1}{2s}\int\overline{N}_{1}\;d\Phi_3,
\end{align}
where $\Phi_3$ is the three-body phase space, bar means averaging
the spins over the initial states and summing up the spins over the
final states. The short-distance part $\bar{N}_0$ and $\bar{N}_1$
defined in Eq.[\ref{eqN0},\ref{eqN1}] can be calculated
perturbatively. The numerical values of the two long-distance matrix
elements $\langle 0|\mathcal{O}_{1}^{\psi}({}^{3}S_{1})|0\rangle$
and $\langle 0|\mathcal{P}_{1}^{\psi}({}^{3}S_{1})|0\rangle$ will be
estimated by nonperturbative methods. In the nonrelativistic
limit, $\langle 0|\mathcal{O}_{1}^{\psi}({}^{3}S_{1})|0\rangle$ can
be related to the nonrelativistic bound state wave function at the
origin. And in NRQCD effective theory, $\langle
0|\mathcal{P}_{1}^{\psi}({}^{3}S_{1})|0\rangle$ also can be
calculated with the help of potential model\cite{Bodwin:2006dn}.
Here we determine them with an alternate way by extracting them from
the decays of $J/\psi$ into light hadrons and into $e^{+}e^{-}$.

\section{Relativistic Corrections to $e^{+}e^{-}\to J/\psi+gg$}

\subsection{Short-Distance Coefficients}
\begin{figure}
\begin{center}
\includegraphics[scale=0.8]{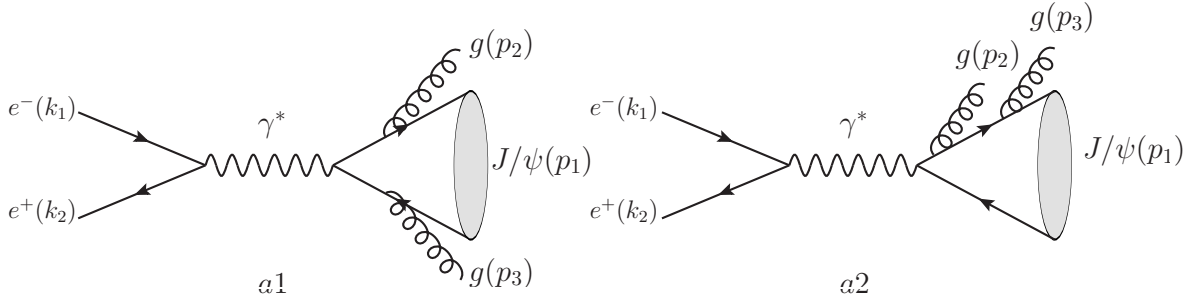}
\caption{The typical two among six Feynman diagrams in the process
$e^{+}e^{-}\to J/\psi+gg$ at leading-order in QCD. }
\end{center}
\end{figure}

At LO in QCD, there are six Feynman diagrams in the process
$e^{+}e^{-}\to J/\psi+gg$, and the typical  two are shown in
Fig.[1]. The full QCD amplitude for this process can be expanded in
terms of the quark relative momentum $q^{\alpha}_{\psi}$:
\begin{eqnarray}
M(e^{+}e^{-}\rightarrow
(C\overline{C})_{{}^{3}S_{1}}(P_{J/\psi})+gg)=
(\frac{m_{c}}{E})^{1/2}A(q_{\psi})=
\nonumber\\
(\frac{m_{c}}{E})^{1/2}(A(0)+q_{\psi}^{\alpha}\frac{\partial
A}{\partial q_{\psi}^{\alpha}}\Big{|}_{q_{\psi}=0}+
\frac{1}{2}q_{\psi}^{\alpha}q_{\psi}^{\beta}\frac{\partial^{2}
A}{\partial q_{\psi}^{\alpha}\partial
q_{\psi}^{\beta}}\Big{|}_{q_{\psi}=0}+\dots),
\end{eqnarray}
where
\begin{eqnarray}
A(q_{\psi})=\sum_{\lambda_{1}\lambda_{2}}\sum_{ij}
\langle\frac{1}{2},\lambda_{1},
\frac{1}{2},\lambda_{2}|1,S_{z}\rangle
\langle3,i;\overline{3},j|1\rangle A(e^{+}e^{-}\rightarrow
c_{\lambda_1,i}(\frac{P}{2}+q_{\psi})\bar{c}_{\lambda_2,j}(\frac{P}{2}-q_{\psi})+
gg),\nonumber\\
\end{eqnarray}
where $\langle 3,i;\bar{3},j|1\rangle =\delta_{ij}/\sqrt{N_c}$ is
the color-SU(3) Clebsch-Gordon coefficient for a $c\bar{c}$ pair
projecting onto a color-singlet state. With the help of
Eq.[\ref{eq10b}], we can express $A(q_{\psi})$ in a covariant form.
The factor $\displaystyle(\frac{m_{c}}{E})^{1/2}$ comes from the
relativistic normalization of the $c\overline{c}$ state, and
$E=\sqrt{m_{c}^2+\mathbf{q_{\psi}}^2}$.

For the $S$ wave state the odd-power terms of $q_{J/\psi}$ vanish
and $q^{\alpha}q^{\beta}=\frac{\mathbf{q}^{2}}{3}(-g^{\alpha
\beta}+\frac{P^{\alpha}P^{\beta}}{P^{2}})=\frac{\mathbf{q}^{2}}{3}\Pi^{\alpha\beta}$,
where $P^2=4E^2, P\cdot q=0$. Then at LO of $v^{2}$ we
have\footnote{We do not expand the three-body phase space by $v_c^2$
and assume $M_{J/\psi}=2m_c$.}
\begin{eqnarray}
|M|^{2}=\frac{m_{c}}{E_{1}}A(0)A^{\ast}(0)+
\frac{1}{2}q_{\psi}^{\alpha}q_{\psi}^{\beta}A_{\alpha\beta}A^{\ast}(0)+
\frac{1}{2}q_{\psi}^{\alpha}q_{\psi}^{\beta}A_{\alpha\beta}^{\ast}A(0),
\end{eqnarray}
where $A_{\alpha\beta}=\frac{\partial^{2}A}{\partial
q^{\alpha}\partial q^{\beta}}$, and
$A_{\alpha\beta}^{\ast}=\frac{\partial^{2}A^{\ast}}{\partial
q^{\alpha}\partial q^{\beta}}$. According to the spinor projection
method the short-distance part of the $v_c^{0}$ part is
\begin{equation}\label{eqN0}
\overline{N}_{0}=\frac{1}{2N_{c}m_{c}}(A(0)A^{\ast}(0))
\Big{|}_{\mathbf{q}_{J/\psi}^{2}=0},
\end{equation}
and the corresponding $v_c^2$ part is
\begin{eqnarray}\label{eqN1}
\overline{N}_{1}=\frac{1}{2N_{c}m_{c}}(\frac{\partial
(\frac{m_{c}}{E}A(0)A^{\ast}(0))}{\partial
(\mathbf{q}_{\psi}^{2})}\Big{|}_{\mathbf{q}_{J/\psi}^{2}=0}
+\frac{1}{6}\Pi_{\alpha\beta}(A_{\psi}^{\alpha\beta}A^{\ast}(0)+
A(0)A_{\psi}^{\ast\alpha\beta})
\Big{|}_{\mathbf{q}_{J/\psi}^{2}=0}).
\end{eqnarray}

We introduce the dimensionless variables $z_{i}=2E_{i}/\sqrt{s},
\overrightarrow{q_{i}}=2\overrightarrow{p_{i}}/\sqrt{s},
x_{i}=\cos\theta_{i}$ and $\delta=4m_c/\sqrt{s}$ to describe the
$e^{-}(k_1)+e^{+}(k_2)\to J/\psi(p_1)+g(p_2)+g({p_3})$ process. Here
$\sqrt{s}$ is the total energy in the center of mass frame,
$p_{1}^{\mu}$, $p_{2}^{\mu}$, $p_{3}^{\mu}$ are the four-momenta of
the final state $J/\psi$ and the two gluons respectively, and
$\theta_{i}$ is the angle between state $i$ and the electron. The
scalar products between the momenta can be expressed as
\begin{align}
&k_{1}\cdot p_{1}=\frac{s}{4}(z_{1}-q_{1}x_{1});\; k_{2}\cdot
p_{1}=\frac{s}{4}(z_{1}+q_{1}x_{1});\; p_{2}\cdot
p_{3}=\frac{s}{8}(4-4z_{1}+\delta^2); \nonumber\\&k_{1}\cdot
p_{2}=\frac{s}{4}(z_{2}-(q_{-}x_{-}-q_{1}x_{1})/2);\; k_{2}\cdot
p_{2}=\frac{s}{4}(z_{2}+(q_{-}x_{-}-q_{1}x_{1})/2);\;
\nonumber\\&k_{1}\cdot
p_{3}=\frac{s}{4}(z_{3}+(q_{-}x_{-}+q_{1}x_{1})/2);\; k_{2}\cdot
p_{3}=\frac{s}{4}(z_{3}-(q_{-}x_{-}+q_{1}x_{1})/2);\;
\nonumber\\&p_{1}\cdot p_{2}=\frac{s}{8}(4-\delta^{2}-4z_{3});\;
p_{1}\cdot p_{3}=\frac{s}{8}(4-\delta^{2}-4z_{2});\;k_{1}\cdot
k_{2}=\frac{s}{2};
\end{align}
where $z_{-}=z_{2}-z_{3}$, $q_{-}=|\vec{q}_{2}-\vec{q}_{3}|
=\sqrt{4-4z_{1}+\delta^{2}+z_{-}^{2}}$,
$q_{1}=|\vec{q}_{1}|=\sqrt{z_{1}-\delta^2}$, $x_{-}=\cos\theta_{-}$,
and $\theta_{-}$ is the angle between $\vec{q}_{-}$ and the
electron. And the three-body phase space is then given by
\begin{align}
d\Phi_{3}&=(2\pi)^{4} \;
\delta^{4}(k_1+k_2-p_1-p_2-p_3)\prod_{i=1}^{3}\frac{d^{3}p_{i}}{2E_{i}}
\nonumber\\&=\frac{s}{32(2\pi)^4}\frac{dz_{1}dx_{1}dz_{-}dw}{\sqrt{(1-K^{2})(1-x_{1}^{2})-w^2}},
\end{align}
where
\begin{subequations}
\begin{equation}
K=\frac{z_{-}(2-z_{1})}{q_{1}q_{-}},
\end{equation}
\begin{equation}
w=x_{-}+Kx_{1}.
\end{equation}
\end{subequations}
The ranges of those integral variables are
\begin{subequations}
\begin{equation}
\delta\leq z_{1}\leq 1+\frac{\delta^2}{4},
\end{equation}
\begin{equation}
-1\leq x_{1} \leq1,
\end{equation}
\begin{equation}
-\sqrt{(z_{1}^{2}-\delta^{2})} \leq
z_{-}\leq\sqrt{(z_{1}^{2}-\delta^{2})},
\end{equation}
\begin{equation}
-\sqrt{(1-K^{2})(1-x_{1}^{2})}\leq w
\leq\sqrt{(1-K^{2})(1-x_{1}^{2})}.
\end{equation}
\end{subequations}

\subsection{Long-Distance Matrix Elements}
The color-singlet production matrix elements can be related to the
decay matrix elements in the vacuum saturation
approximation\cite{Bodwin:1994jh} and the errors are of $v_c^4$
order. Since in this $e^{+}e^{-}\to J/\psi+gg$ process, there are
two NRQCD matrix elements accurate to order $v_c^{2}$, i.e.,
$\langle0|\mathcal{O}_{1}^{\psi}({}^{3}S_{1})|0\rangle$ and
$\langle0|\mathcal{P}_{1}^{\psi}({}^{3}S_{1})|0\rangle$, we can
determine their values by fitting $J/\psi$ decays into $e^{+}e^{-}$
and   into light hadrons (LH). The theoretical results at NLO in
$\alpha_{s}$ and $v^{2}$ for $J/\psi \rightarrow e^{+}e^{-}$ and
$J/\psi \rightarrow LH$ \footnote{We do not include the
electromagnetic process $J/\psi\rightarrow \gamma^{*}\rightarrow
LH.$ } are \cite{Bodwin:2002hg}
\begin{subequations}\label{ThDecay}
\begin{equation}
\Gamma(J/\psi\rightarrow
e^{+}e^{-})=\frac{2e_{c}^{2}\pi\alpha^{2}}{3}
\Big((1-\frac{16\alpha_{s}}{3\pi})\frac{\langle0|\mathcal{O}_{1}^{\psi}({}^{3}S_{1})|0\rangle/3}{m_{c}^{2}}-
\frac{4}{3}\frac{\langle0|\mathcal{P}_{1}^{\psi}({}^{3}S_{1})|0\rangle/3}{m_{c}^{4}}\Big),
\end{equation}
\begin{equation}
\Gamma(J/\psi\rightarrow
LH)=(\frac{20\alpha_{s}^{3}}{243}(\pi^{2}-9))
\Big((1-2.55\frac{\alpha_{s}}{\pi})\frac{\langle0|\mathcal{O}_{1}^{\psi}({}^{3}S_{1})|0\rangle/3}{m_{c}^{2}}-
\frac{19\pi^{2}-132}{12\pi^{2}-108}
\frac{\langle0|\mathcal{P}_{1}^{\psi}({}^{3}S_{1})|0\rangle/3}{m_{c}^{4}}\Big).
\end{equation}
\end{subequations}
And the central values of the experimental results
are\cite{Amsler:2008zzb}
\begin{equation}
\Gamma(J/\psi\rightarrow e^{+}e^{-})=5.54\mathrm{keV},\;
\Gamma(J/\psi\rightarrow LH)=69.2\mathrm{keV}
\end{equation}
Solving these equations at LO of $\alpha_{s}$ (QCD
radiative corrections not included), we obtain
\begin{equation}\label{eqM1}
\frac{\langle0|\mathcal{O}_{1}^{\psi}({}^{3}S_{1})|0\rangle}{3}=0.294\mathrm{GeV^{3}},\;
\frac{\langle0|\mathcal{P}_{1}^{\psi}({}^{3}S_{1})|0\rangle}{3m_{c}^{2}}=0.320
\times10^{-1}\mathrm{GeV^{3}},
\end{equation}
for $m_{c}=$1.5GeV and $\alpha_{s}=0.26$. Fixing $\alpha_s=0.26$, we
can express the matrix elements as functions of $m_c$, which are
\begin{eqnarray}\label{eqM2}
\frac{\langle0|\mathcal{O}_{1}^{\psi}({}^{3}S_{1})|0\rangle}{3}=0.131m_{c}^{2},\;
\frac{\langle0|\mathcal{P}_{1}^{\psi}({}^{3}S_{1})|0\rangle}{3m_{c}^{2}}=0.142\times10^{-1}m_{c}^{2}.
\end{eqnarray}
If we include the QCD NLO radiative corrections in
Eq.[\ref{ThDecay}], for $m_{c}=$1.5Gev and $\alpha_{s}=0.26$, we get
\begin{equation}\label{eqM3}
\frac{\langle0|\mathcal{O}_{1}^{\psi}({}^{3}S_{1})|0\rangle}{3}=0.572\mathrm{GeV^{3}},\;
\frac{\langle0|\mathcal{P}_{1}^{\psi}({}^{3}S_{1})|0\rangle}{3m_{c}^{2}}=0.512
\times10^{-1}\mathrm{GeV^{3}}.
\end{equation}
The corresponding $m_c$ dependence of the matrix elements are
\begin{eqnarray}\label{eqM4}
\frac{\langle0|\mathcal{O}_{1}^{\psi}({}^{3}S_{1})|0\rangle}{3}=0.254m_{c}^{2},\;
\frac{\langle0|\mathcal{P}_{1}^{\psi}({}^{3}S_{1})|0\rangle}{3m_{c}^{2}}=0.228\times10^{-1}m_{c}^{2}.
\end{eqnarray}

In Ref.\cite{Gremm:1997dq}, the authors relate
$\langle0|\mathcal{P}_{1}^{\psi}({}^{3}S_{1})|0\rangle$ to
$\langle0|\mathcal{O}_{1}^{\psi}({}^{3}S_{1})|0\rangle$ using the
NRQCD equation of motion . In Ref.\cite{Bodwin:2006dn},
$\langle0|\mathcal{P}_{1}^{\psi}({}^{3}S_{1})|0\rangle$ is
calculated by combining NRQCD and the potential model.
Alternately, we estimate the production matrix elements by using the
experimentally observed charmonium decay rates. As argued
in\cite{Bodwin:1994jh}, the differences between the color-singlet
production and decay matrix elements are of order $v^4$. So our
method should be valid at order $v^{2}$, and our numerical results
of the production matrix elements are adequate, if the higher order
QCD and $v^{2}$ corrections to the decay rates are small and the
uncertainties of the experimental data are not large. If we define
$O(\langle
v^2\rangle)=\frac{\langle0|\mathcal{P}_{1}^{\psi}({}^{3}S_{1})|0\rangle}
{m_c^2\langle0|\mathcal{O}_{1}^{\psi}({}^{3}S_{1})|0\rangle}$, we
find the value of $O(\langle v^2\rangle)$ to be about 0.1 from
Eq.(\ref{eqM1}) or Eq.(\ref{eqM3}), which is about $2\sim3$ times
smaller than the result calculated within a potential
model\cite{Bodwin:2006dn,Bodwin:2007fz} or within the QCD sum
rules\cite{Braguta:2006wr}. Note that, differing from theirs, our
matrix elements are extracted by fitting experimental data, which
depend on the chosen processes and experimental errors. Moreover,
the short-distance coefficient of the $v^2$ correction term in
Eq.(\ref{ThDecay}b) for $J/\psi\to \mathrm{LH}$ is quite large,
implying that the corresponding long-distance matrix element could
be rather small. These of course will have uncertainties, compared
with other methods for calculating the long-distance matrix
elements. Nevertheless, our method using experimental data to
extract the matrix elements, provides an independent and
self-consistent way to determine the matrix elements. So in this
work we will use these experimentally extracted long-distance matrix
elements to give numerical predictions.

\subsection{Numerical Result}

The expression of $N_{0}$ and $N_{1}$ are too complicated to be
shown here and we only give the analytical expression of the
differential cross section. With $J^{\mathrm{PC}}$ conservation and gauge
invariance, the general form of the differential cross section of
unpolarized $J/\psi$ production in $e^{+}e^{-}$ annihilation can be
expressed as
\begin{equation}\label{Salpha}
\frac{d^{2}\sigma}{dz_{1}dx_{1}}(e^{+}e^{-}\rightarrow
\gamma^{\ast}\rightarrow\psi+gg)=S_0(z_{1})(1+\alpha_0(z_{1})
x_{1}^{2})+ S_v(z_{1})(1+\alpha_v(z_{1}) x_{1}^{2}),
\end{equation}
where the first term on the right-hand side of Eq.[\ref{Salpha}] is
the LO result in $v_c$, and the second one is the relativistic
correction term. Our results of $S_0(z_{1})$ and $\alpha_0(z_{1})$
are in consistency with those in \cite{Cho:1996cg}. And the
expressions of $S_v(z_{1})$ and $S_v(z_{1})\alpha_v(z_{1})$ are too
complicated and they will not be given here.

Setting $\alpha_s=0.26$, $\sqrt{s}=10.6\mathrm{GeV}$ and
$m_c=1.5\mathrm{GeV}$ and integrating out $x_1$ and $z_1$
numerically, we find when choosing the values of the matrix elements
in Eq.[\ref{eqM1}], the LO result is
\begin{equation}
\sigma_{LO}(e^{+}+e^{-}\to J/\psi+gg)=202\;\mathrm{fb}.
\end{equation}
and the relativistic correction is $55\;\mathrm{fb}$, which gives
about $27\%$ enhancement, then resulting in the NLO result in
$v_c^2$:
\begin{equation}\label{relan}
\sigma_{\mathrm{NLO}(v_c^2)}(e^{+}+e^{-}\to J/\psi+gg)=257\;\mathrm{fb}.
\end{equation}
If we choose the values of the matrix elements in Eq.[\ref{eqM3}],
the NLO result in $v_c^2$ becomes
\begin{equation}\label{rela}
\sigma_{\mathrm{NLO}(v_c^2)}(e^{+}+e^{-}\to J/\psi+gg)=480\;\mathrm{fb},
\end{equation}
and the relativistic correction enhancement is $1.22$. When the
charm-quark mass varies from $1.4\mathrm{GeV}$ to $1.6\mathrm{GeV}$,
the LO and NLO cross sections as function of $m_c$ are shown in
Fig.[2] with  the long-distance matrix elements, respectively, in
Eq.(\ref{eqM2}) and Eq.(\ref{eqM4}). As mentioned above, we do not
expand the three-body phase space by $v_c^2$ and assume $M_{J/\psi}=2m_c$
for simplicity. Instead, if we used the physical mass of
$M_{J/\psi}=3.097\mathrm{GeV}$, which includes $v_c^2$ kinematic and
binding energy corrections, when doing the phase space integrals,
the results given in Eq.(\ref{relan}) and Eq.(\ref{rela}) turned to
be 253fb and 475fb, respectively. Comparing those results, one can
see that the differences due to replacing
$M_{J/\psi}=3.097\mathrm{GeV}$ with $M_{J/\psi}=2m_c$ in the phase
space integration can be neglected.

To be consistent, when we calculate the production cross section
including both the $O(v^2)$ relativistic correction  and the
$O(\alpha_s)$ radiative correction, we should first use the
relativistic correction results obtained by adopting
Eqs.(\ref{eqM3},\ref{eqM4}), where the $O(\alpha_s)$ radiative
corrections to the matrix elements are included, and then further
include the $O(\alpha_s)$ radiative corrections to the production
short-distance coefficients. This will be discussed in next section.
\begin{figure}
\begin{center}
\includegraphics[scale=0.7]{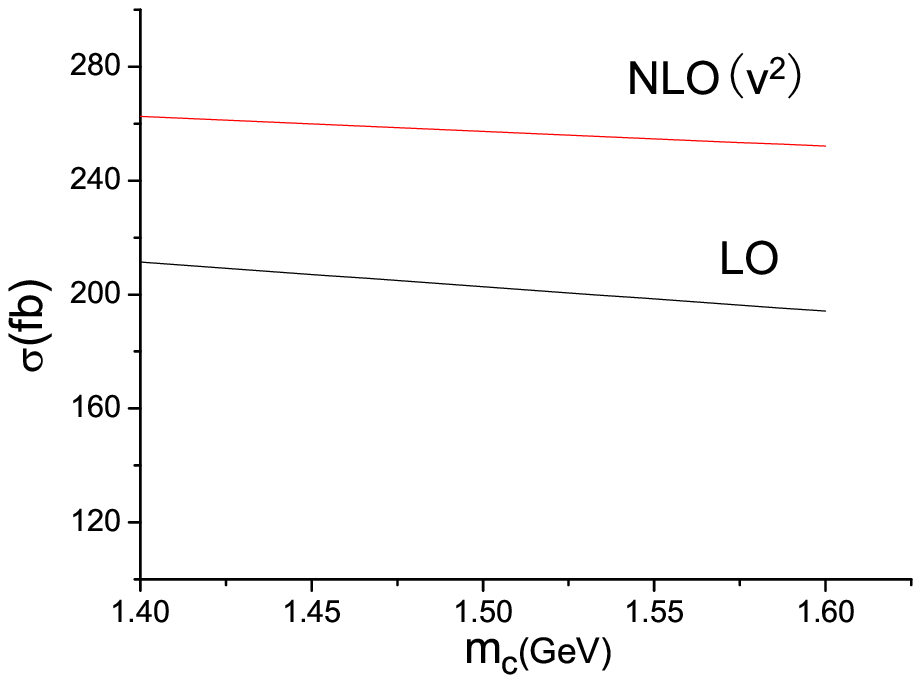}
\includegraphics[scale=0.7]{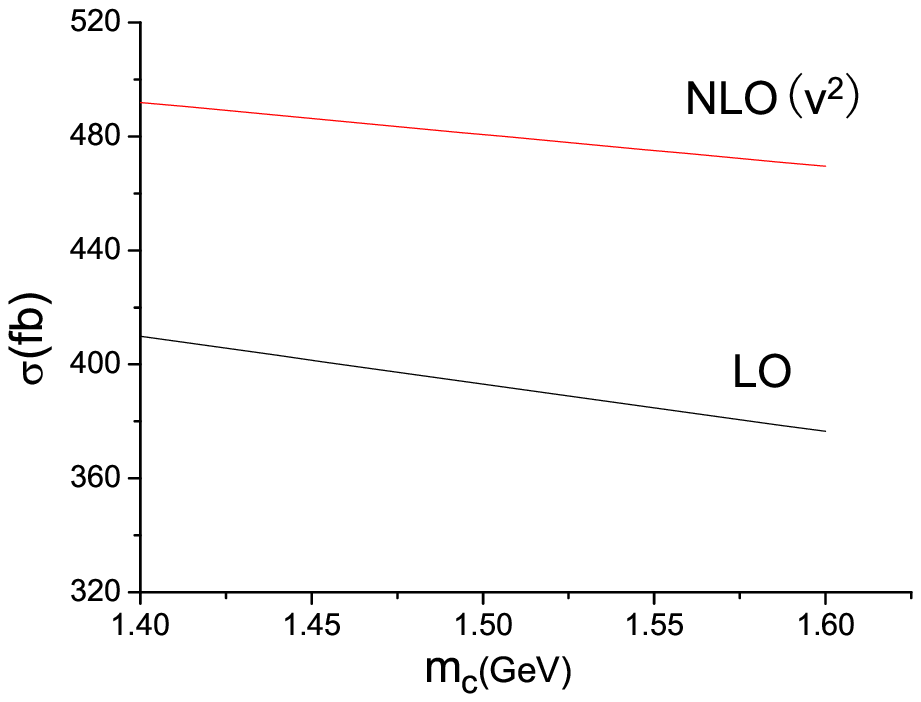}
\caption{Relativistic corrections to $\sigma(e^{+}e^{-}\to
J/\psi+gg)$ as functions of $m_c$ with long-distance matrix elements
determined from Eq.(\ref{eqM2}) (left-panel) and Eq.(\ref{eqM4}) (right-panel). In
each figure the lower curve is the LO result and the upper one is
the result including $v_c^2$ corrections.}
\end{center}
\end{figure}

\section{Discussion and Summary}

In above the order $v_c^2$ relativistic effect is considered for the
color-singlet $e^{+}e^{-}\to J/\psi+gg$ process in the framework of
NRQCD. We calculate the short-distance coefficients perturbatively
and find the ratio of the coefficient of
$\langle0|\mathcal{P}_{1}^{\psi}({}^{3}S_{1})|0\rangle/m_c^2$ to
that of $\langle0|\mathcal{O}_{1}^{\psi}({}^{3}S_{1})|0\rangle$ is
about $2.5$. Denoting the ratio of the long-distance matrix elements
by $\langle0|\mathcal{P}_{1}^{\psi}({}^{3}S_{1})|0\rangle/
(m_c^2\langle0|\mathcal{O}_{1}^{\psi}({}^{3}S_{1})|0\rangle)=
\langle v_c^2\rangle$\cite{Bodwin:2006dn}, then the enhancement of
the relativistic correction can be expressed by $2.5\langle
v_c^2\rangle$. Using the matrix elements given in Eq.[\ref{eqM1}]
and Eq.[\ref{eqM3}] as inputs, we predict the enhancements of
relativistic corrections are $22\%$ and $27\%$, respectively, which
are as important as the NLO QCD corrections\cite{Ma:2008gq}. It can
also be found that when including the relativistic corrections the
$m_c$ dependence is also improved a little. If we determine the
matrix element from the $J/\psi\to e^{+}e^{-}$ process in
Eq.[\ref{ThDecay}a] without including the relativistic and QCD
corrections, we get
$\frac{\langle0|\mathcal{O}_{1}^{\psi}({}^{3}S_{1})|0\rangle}{3}=0.251\mathrm{GeV^{3}}$
and the LO result of $e^{+}e^{-}\to J/\psi+gg$ is only about
$173\mathrm{fb}$. Then we can see that the relativistic corrections
can enhance both the short-distance coefficients and the
long-distance matrix elements.

Next, we further include the $O(\alpha_s)$ radiative corrections to
the production short-distance coefficients. In Ref.\cite{Ma:2008gq},
the authors obtain that with $\alpha_s(2m_c)=0.259$ the K-factor of
NLO QCD result to LO QCD result is $1.20$. Then using the matrix
elements in Eq.[\ref{eqM3}], which also include the NLO QCD
corrections in decay processes, we find that after including the QCD
corrections\cite{Ma:2008gq} the combined cross section
$\sigma_{\mathrm{NLO}(v_c^2,\alpha_s)}$ for $m_c=1.5\mathrm{GeV}$ is
\begin{equation}\label{Fresult}
\sigma_{\mathrm{NLO}(v_c^2,\alpha_s)}(e^{ +}e^{-}\to J/\psi+gg)\simeq
480/1.22\times(1+0.22+0.20)\simeq 560\mathrm{fb},
\end{equation}
where on the right-hand side of Eq.(\ref{Fresult}) the number 480 fb
comes from Eq.(\ref{rela}), and in the summation the enhancement
factor 0.22 is due to relativistic correction, while the enhancement
factor 0.20 due to QCD radiative correction. Note that all the above
contributions come from the color-singlet part. The LO color-octet
contribution of $e^{+}e^{-}\to J/\psi+g$ can be estimated as large
as $0.27\mathrm{pb}$\cite{Wang:2003fw}, but this apparently depends
on the chosen values of the color-octet matrix elements. If using
this estimate\cite{Wang:2003fw} for the color-octet contribution,
then the prediction of NRQCD for the
$J/\psi+X_{\mathrm{non-c\bar{c}}}$ cross section at $B$-factories
would become $0.83\mathrm{pb}$, which is almost twice as large as
the measured central value $0.43\mathrm{pb}$ by Belle\cite{:2009nj}.

In fact, from Eq.(\ref{Fresult}) we see that after including the QCD
and relativistic corrections the color-singlet contribution alone
has saturated the measured value of
$J/\psi+X_{\mathrm{non-c\bar{c}}}$ cross section, and thus there
seems no need for the color-octet contribution. However, we must pay
attention to possible uncertainties before we can draw a firm
conclusion. First, on the experimental side, there is a large
uncertainty of Belle's result in the $p_{j/\psi}<2.0\mathrm{GeV}$
region, and furthermore the total cross sections of inclusive
$J/\psi$ production measured by BaBar, Belle, and CLEO are not
consistent. Second, on the theoretical side, we should take into
account the uncertainty due to the choice of renormalization scale
$\mu$ in the calculation of the short-distance coefficients. In
doing the latter, we choose the largest value $\mu=\sqrt{s}/2$ and
$m_c=1.5\mathrm{GeV}$, and find  the K-factor of NLO QCD correction
to be 1.48 (see Ref.\cite{Ma:2008gq}). We then combine the
relativistic correction with the QCD radiative correction, and find
that at NLO in $v_c^2$ and $\alpha_s$ the total cross section of
direct $J/\psi$ production becomes
\begin{equation}\label{muFresult} \sigma_{\mathrm{NLO}(v_c^2,\alpha_s)}(e^{+}e^{-}\to
J/\psi+gg)\simeq437\mathrm{fb}
\end{equation}
for $\mu=\sqrt{s}/2$ and $m_c=1.5\mathrm{GeV}$. We see that although
the cross section is decreased as compared with that for $\mu=2m_c$
and $m_c=1.5\mathrm{GeV}$, the predicted cross section is still a
little larger than the central value of the latest Belle result.

Moreover, the cross sections obtained in Eq.(\ref{Fresult}) and
Eq.(\ref{muFresult}) are the direct $J/\psi$ production rates, not
including the feed-down contribution from higher charmonium states.
If the feed-down contribution is included, the prompt $J/\psi$
production cross section, which is the measured value by Belle, will
be further enhanced by a factor of about 1.3 (see
Refs.\cite{Ma:2008gq,Gong:2009kp} for discussions on the feed-down
contribution). Therefore, the theoretical cross section for the
$J/\psi$ prompt production calculated at NLO in $\alpha_s$ and $v^2$
in NRQCD will exceed or saturate the latest value in
Eq.(\ref{noncc}) observed by Belle\cite{:2009nj}, despite of
theoretical uncertainties related to the choice of input parameters,
e.g., $\mu$, $m_c$, and the color-singlet matrix elements.

In summary, we find the $O(v^2)$ relativistic correction to enhance
the cross section of  $J/\psi$ production in the color-singlet
process $e^{+}e^{-}\to J/\psi+gg$ by a factor of 20-30\%, which is
comparable to the enhancement caused by the $O(\alpha_s)$ radiative
correction\cite{Ma:2008gq,Gong:2009kp}. As the consequence of
including both the $O(\alpha_s)$ radiative correction and the
$O(v^2)$ relativistic correction, the color-singlet contribution to
$e^{+}e^{-}\to J/\psi+gg$ has saturated the latest observed cross
section by Belle for $e^{+}e^{-}\to
J/\psi+X_{\mathrm{non-c\bar{c}}}$ at $B$-factories, thus leaving
little room for the color-octet contribution. This gives a very
stringent constraint on the color-octet contribution, and may imply
that the values of color-octet matrix elements are much smaller than
expected earlier by using the naive velocity scaling rules or
extracted from fitting experimental data with the LO results. To
reduce the theoretical uncertainties, further investigations for the
higher order (both in $\alpha_s$ and $v^2$) corrections are needed.
Moreover, comparisons between various experimental measurements and
theoretical predictions are certainly helpful to clarify this
important issue concerning the color-octet mechanism.

\section{Acknowledgement}
We thank Yan-Qing Ma for useful discussions. This work was supported
by the National Natural Science Foundation of China (No. 10675003, No.
10721063) and the Ministry of Science and Technology of China
(2009CB825200). Zhi-Guo He is currently supported by the CPAN08-PD14
contract of the CSD2007-00042 Consolider-Ingenio 2010 program, and
by the FPA2007-66665-C02-01/ project (Spain).


\begin{thebibliography}{99}

\bibitem{Bodwin:1994jh}
  G.~T.~Bodwin, E.~Braaten and G.~P.~Lepage,
  Phys.\ Rev.\  D {\bf 51}, 1125 (1995)
  [Erratum-ibid.\  D {\bf 55}, 5853 (1997)]
  [arXiv:hep-ph/9407339].


\bibitem{Brambilla:2004wf}
  N.~Brambilla {\it et al.}  [Quarkonium Working Group],
  [arXiv:hep-ph/0412158].


\bibitem{Aubert:2001pd}
  B.~Aubert {\it et al.}  [BABAR Collaboration],
  Phys.\ Rev.\ Lett.\  {\bf 87}, 162002 (2001)
  [arXiv:hep-ex/0106044].


\bibitem{Abe:2001za}
  K.~Abe {\it et al.}  [BELLE Collaboration],
  Phys.\ Rev.\ Lett.\  {\bf 88}, 052001 (2002)
  [arXiv:hep-ex/0110012].


\bibitem{Briere:2004ug}
  R.~A.~Briere {\it et al.}  [CLEO Collaboration],
  Phys.\ Rev.\  D {\bf 70}, 072001 (2004)
  [arXiv:hep-ex/0407030].


\bibitem{Driesen:1993us}
  V.~M.~Driesen, J.~H.~Kuhn and E.~Mirkes,
  Phys.\ Rev.\  D {\bf 49}, 3197 (1994).


\bibitem{Cho:1996cg}
  P.~L.~Cho and A.~K.~Leibovich,
  Phys.\ Rev.\  D {\bf 54}, 6690 (1996)
  [arXiv:hep-ph/9606229].


\bibitem{Baek:1998yf}
  S.~Baek, P.~Ko, J.~Lee and H.~S.~Song,
  J.\ Korean Phys.\ Soc.\  {\bf 33}, 97 (1998)
  [arXiv:hep-ph/9804455].

\bibitem{Yuan:1996ep}
  F.~Yuan, C.~F.~Qiao and K.~T.~Chao,
  Phys.\ Rev.\  D {\bf 56}, 321 (1997)
  [arXiv:hep-ph/9703438].




\bibitem{Braaten:1995ez}
  E.~Braaten and Y.~Q.~Chen,
  Phys.\ Rev.\ Lett.\  {\bf 76}, 730 (1996)
  [arXiv:hep-ph/9508373].



\bibitem{Wang:2003fw}
  J.~X.~Wang,
  arXiv:hep-ph/0311292.


\bibitem{Fleming:2003gt}
  S.~Fleming, A.~K.~Leibovich and T.~Mehen,
  Phys.\ Rev.\  D {\bf 68}, 094011 (2003)
  [arXiv:hep-ph/0306139].


\bibitem{Abe:2002rb}
  K.~Abe {\it et al.}  [Belle Collaboration],
  Phys.\ Rev.\ Lett.\  {\bf 89}, 142001 (2002)
  [arXiv:hep-ex/0205104].


\bibitem{:2009nj}
  P.~Pakhlov {\it et al.}  [Belle Collaboration],
  Phys.\ Rev.\  D {\bf 79}, 071101 (2009)
  [arXiv:0901.2775 [hep-ex]].


\bibitem{Liu:2003jj}
  K.~Y.~Liu, Z.~G.~He and K.~T.~Chao,
  Phys.\ Rev.\  D {\bf 69}, 094027 (2004)
  [arXiv:hep-ph/0301218].


\bibitem{Zhang:2006ay}
  Y.~J.~Zhang and K.~T.~Chao,
  Phys.\ Rev.\ Lett.\  {\bf 98}, 092003 (2007)
  [arXiv:hep-ph/0611086].



\bibitem{Gong:2009ng}
  B.~Gong and J.~X.~Wang,
  Phys.\ Rev.\  D {\bf 80}, 054015 (2009)
  [arXiv:0904.1103 [hep-ph]].




\bibitem{Ma:2008gq}
  Y.~Q.~Ma, Y.~J.~Zhang and K.~T.~Chao,
  Phys.\ Rev.\ Lett.\  {\bf 102}, 162002 (2009)
  [arXiv:0812.5106 [hep-ph]].
\bibitem{Gong:2009kp}
  B.~Gong and J.~X.~Wang,
  Phys.\ Rev.\ Lett.\  {\bf 102}, 162003 (2009)
  [arXiv:0901.0117 [hep-ph]];


\bibitem{He:2007te}
  Z.~G.~He, Y.~Fan and K.~T.~Chao,
  Phys.\ Rev.\  D {\bf 75}, 074011 (2007)
  [arXiv:hep-ph/0702239].


\bibitem{Pakhlov:2004au}
  P.~Pakhlov  [Belle Collaboration],
  arXiv:hep-ex/0412041.

\bibitem{Aubert:2005tj}
  B.~Aubert {\it et al.}  [BABAR Collaboration],
  Phys.\ Rev.\  D {\bf 72}, 031101 (2005)
  [arXiv:hep-ex/0506062].


\bibitem{Braaten:2002fi}
  E.~Braaten and J.~Lee,
  Phys.\ Rev.\  D {\bf 67}, 054007 (2003)
  [Erratum-ibid.\  D {\bf 72}, 099901 (2005)]
  [arXiv:hep-ph/0211085].
  K.~Y.~Liu, Z.~G.~He and K.~T.~Chao,
  Phys.\ Lett.\  B {\bf 557}, 45 (2003)
  [arXiv:hep-ph/0211181]; Phys. Rev. D 77, 014002
(2008)[arXiv:hep-ph/0408141];
  K.~Hagiwara, E.~Kou and C.~F.~Qiao,
  Phys.\ Lett.\  B {\bf 570}, 39 (2003)
  [arXiv:hep-ph/0305102].


\bibitem{Zhang:2005cha}
  Y.~J.~Zhang, Y.~J.~Gao and K.~T.~Chao,
  Phys.\ Rev.\ Lett.\  {\bf 96}, 092001 (2006)
  [arXiv:hep-ph/0506076].

  B.~Gong and J.~X.~Wang,
  Phys.\ Rev.\  D {\bf 77}, 054028 (2008)
  [arXiv:0712.4220 [hep-ph]].

\bibitem{Bodwin:2006ke}
  G.~T.~Bodwin, D.~Kang, T.~Kim, J.~Lee and C.~Yu,
  AIP Conf.\ Proc.\  {\bf 892} (2007) 315
  [arXiv:hep-ph/0611002].


\bibitem{Fan:2009zq}
  Y.~Fan, Y.~Q.~Ma and K.~T.~Chao,
  Phys.\ Rev.\  D {\bf 79}, 114009 (2009)
  [arXiv:0904.4025 [hep-ph]].


\bibitem{Kuhn:1979bb}
  J.~H.~Kuhn, J.~Kaplan and E.~G.~O.~Safiani,
  Nucl.\ Phys.\  B {\bf 157}, 125 (1979);
  B.~Guberina, J.~H.~Kuhn, R.~D.~Peccei and R.~Ruckl,
  Nucl.\ Phys.\  B {\bf 174}, 317 (1980);
  E.~L.~Berger and D.~L.~Jones,
  Phys.\ Rev.\  D {\bf 23}, 1521 (1981).


\bibitem{Keung:1982jb}
  W.~Y.~Keung and I.~J.~Muzinich,
  Phys.\ Rev.\  D {\bf 27}, 1518 (1983);


\bibitem{Bodwin:2002hg}
  G.~T.~Bodwin and A.~Petrelli,
  Phys.\ Rev.\  D {\bf 66}, 094011 (2002)
  [arXiv:hep-ph/0205210].


\bibitem{Braaten:1996jt}
  E.~Braaten and Y.~Q.~Chen,
  Phys.\ Rev.\  D {\bf 54}, 3216 (1996)
  [arXiv:hep-ph/9604237].


\bibitem{Bodwin:2006dn}
  G.~T.~Bodwin, D.~Kang and J.~Lee,
  Phys.\ Rev.\  D {\bf 74}, 014014 (2006)
  [arXiv:hep-ph/0603186].


\bibitem{Amsler:2008zzb}
  C.~Amsler {\it et al.}  [Particle Data Group],
  Phys.\ Lett.\  B {\bf 667}, 1 (2008).


\bibitem{Gremm:1997dq}
  M.~Gremm and A.~Kapustin,
  Phys.\ Lett.\  B {\bf 407}, 323 (1997)
  [arXiv:hep-ph/9701353].


\bibitem{Bodwin:2007fz}
  G.~T.~Bodwin, H.~S.~Chung, D.~Kang, J.~Lee and C.~Yu,
  Phys.\ Rev.\  D {\bf 77}, 094017 (2008)
  [arXiv:0710.0994 [hep-ph]].


\bibitem{Braguta:2006wr}
  V.~V.~Braguta, A.~K.~Likhoded and A.~V.~Luchinsky,
  Phys.\ Lett.\  B {\bf 646}, 80 (2007)
  [arXiv:hep-ph/0611021].





\end{thebibliography}
\end{document}